\documentclass{article}
\usepackage{spconf,amsmath,graphicx,xcolor}

\graphicspath{{./Figures/}}

\def\F0{{F$_0$}}

\newcommand\sectioncut{-2.2mm}

\title{EmoCat: Language-agnostic Emotional Voice Conversion}

\name{Bastian Schnell$^{\ddagger}$\sthanks{Work performed while being an intern at Amazon.}, Goeric Huybrechts$^{\dagger}$, Bartek Perz$^{\dagger}$, Thomas Drugman$^{\dagger}$,  Jaime Lorenzo-Trueba$^{\dagger}$}
\address{\begin{tabular}{cc}$^{\ddagger}$ Idiap Research Institute, Martigny, Switzerland\\
		$^{\dagger}$ Amazon, TTS Research, Cambridge, United Kingdom\\
	{\normalsize\texttt{bastian.schnell@idiap.ch, \{huybrech,perzbart,drugman,truebaj\}@amazon.com}}\end{tabular}}

\begin{document}
\ninept
\maketitle

\begin{abstract}
Emotional voice conversion models adapt the emotion in speech without changing the speaker identity or linguistic content. They are less data hungry than text-to-speech models and allow to generate large amounts of emotional data for downstream tasks. 
In this work we propose EmoCat, a language-agnostic emotional voice conversion model. It achieves high-quality emotion conversion in German with less than 45 minutes of German emotional recordings by exploiting large amounts of emotional data in US English. EmoCat is an encoder-decoder model based on CopyCat, a voice conversion system which transfers prosody. We use adversarial training to remove emotion leakage from the encoder to the decoder. The adversarial training is improved by a novel contribution to gradient reversal to truly reverse gradients. This allows to remove only the leaking information and to converge to better optima with higher conversion performance. Evaluations show that Emocat can convert to different emotions but misses on emotion intensity compared to the recordings, especially for very expressive emotions. EmoCat is able to achieve audio quality on par with the recordings for five out of six tested emotion intensities.\footnote{Audio samples will be released after paper acceptance.}
\end{abstract}
\vspace{\sectioncut}
\section{Introduction}
\label{sec:intro}
Neural Text-to-Speech (TTS) has greatly supported the advent of artificial voice assistants like Amazon Alexa, Google Assistant, or Siri. These systems are trained on tens of hours of data  \cite{aggarwal2020using} and produce high-quality speech with close to perfect intelligibility \cite{shen2018natural}. However, their speech is mostly neutral, which prevents natural conversations and closer bounds with the user. Creating voices in more expressive speaking styles usually requires recording similarly large amounts of speech for the desired style. This is very time-consuming and costly. An alternative is the generation of synthetic data to satisfy the high data needs. The conversion of speech is generally assumed to be easier than TTS, thus has lower data needs.

Emotional voice conversion (EVC) is a subfield of voice conversion (VC) which studies the transformation of a source audio signal into a different emotion while maintaining its linguistic content and speaker identity. Techniques applied in EVC are similar to VC and differ mostly in their feature selection \cite{kameoka2018stargan,rizos2020stargan}. EVC techniques working without hand-crafted features are applicable to other speaking styles as well. 
EVC is also applied to other tasks like film dubbing.

In this work, we aim to convert neutral to emotional speech in German. As we have only a limited amount of emotional German data available, we exploit emotional recordings in US English. We propose EmoCat, a language-agnostic EVC model trained jointly on German and US English working directly on mel-spectrograms. Compared to other works we use mel-spectrograms to leverage our high-quality universal vocoder \cite{lorenzo2019towards} to keep a high bar on segmental quality. Our model adapts the CopyCat model \cite{karlapati2020copycat} (which is based on AutoVC \cite{qian2019autovc}) for intra-speaker emotion conversion. CopyCat is a VC model which allows to convert the speech of unseen speakers to a set of target speakers. In contrast to the global speaker identity, emotion is a continuous component of speech. We use adversarial training to explicitly remove emotion leakage from the encoder, which encodes the neutral source spectrogram, to the decoder, which generates the converted emotional spectrogram. We propose a novel improvement to gradient reversal \cite{ganin2015unsupervised} to stabilise its gradients. We further investigate fine-tuning to improve naturalness. In an ablation study, we assess the effectiveness of each of the techniques. The proposed model is able to convert neutral German to two different emotions in three intensities with the support of less than 45 minutes of German emotional data. To the best of our knowledge, no work exists on EVC with multi-lingual data or mel-spectrograms.

\vspace{\sectioncut}
\section{Related work}
\label{sec:related_work}
Emotional voice conversion methods are generally split into two categories: parallel and non-parallel training data.
In the parallel data scenario, the database contains the same utterance spoken by the same speaker in the different target emotions.

In the non-parallel data scenario, the utterances for each emotion differ,meaning that the content can better match the emotion. This allows a wider variety of utterances and also simplifies acting for the voice talents. With the lack of parallel data, a model cannot be trained to do the conversion directly as the ground truth target is not available. The training can only be guided in an unsupervised way. Generative adversarial networks (GAN) and cycle consistency losses are commonly used techniques here \cite{gao2019nonparallel,rizos2020stargan,zhou2020transforming}.

In \cite{gao2019nonparallel} an encoder-decoder structure with a content and style encoder
is used to convert mel-cepstrum (MCEP) extracted by WORLD  \cite{morise2016world} . The model is trained with three losses. First, the cepstrogram is auto-encoded and an L1 reconstruction loss applied.
Second, a semi-cycle consistency L1 loss forces the encoder embeddings to match before and after conversion.
Third, a GAN loss tries to discriminate generated from recorded samples.
\F0 is converted by a linear transform
to match the statistics of the target emotion domain. The band aperiodicities remain unchanged.

StarGAN is used in \cite{rizos2020stargan} on WORLD features with a reconstruction loss, an L1 cycle consistency loss, and a real/fake GAN loss. The model architecture is the same as in the original StarGAN-VC paper \cite{kameoka2018stargan}. An emotion recognition model (a variant of \cite{zhang2019attention}) was trained with the generated samples and evaluations show that its accuracy improved.

CycleGAN has also been used for emotion conversion \cite{zhou2020transforming}. It is trained with three losses: 1) a reconstruction loss, 2) a cycle-consistency loss on a sample converted to another emotion and then back to the source emotion, and 3) the GAN loss for real/fake discrimination. The experiments show that separate CycleGANs for
F0 and MCEP outperform a joint model. \cite{liu2020emotional} follows a very similar approach but uses an additional emotion classification loss and no reconstruction loss.

A different approach is the variational auto-encoding Wasserstein GAN (VAW-GAN) for emotion conversion \cite{zhou2020converting} (originally proposed for VC in \cite{hsu2017voice}). It consists of a variational auto-encoder (VAE) structure where the decoder is conditioned on an emotion embedding. The latent dimension is chosen to be small enough so that it will not contain emotion information. The model is trained with reconstruction loss, standard Kullback-Leibler (KL) divergence on the VAE latent space, and a Wasserstein GAN loss. Instead of using a binary cross-entropy loss for the real/fake prediction of the discriminator, the Wasserstein distance is used.

Our approach is closest to the VAW-GAN in \cite{zhou2020converting} as it employs a similar encoder-decoder structure with a VAE encoder. However, the bottleneck used is temporal and drastically smaller, also we condition the decoder on the linguistic content. In contrast to all related work above, we operate on mel-spectrograms and train with multi-lingual data.

\vspace{\sectioncut}
\section{Model description}
\label{sec:model_description}

\begin{figure}
	\centering
	\includegraphics[width=\columnwidth]{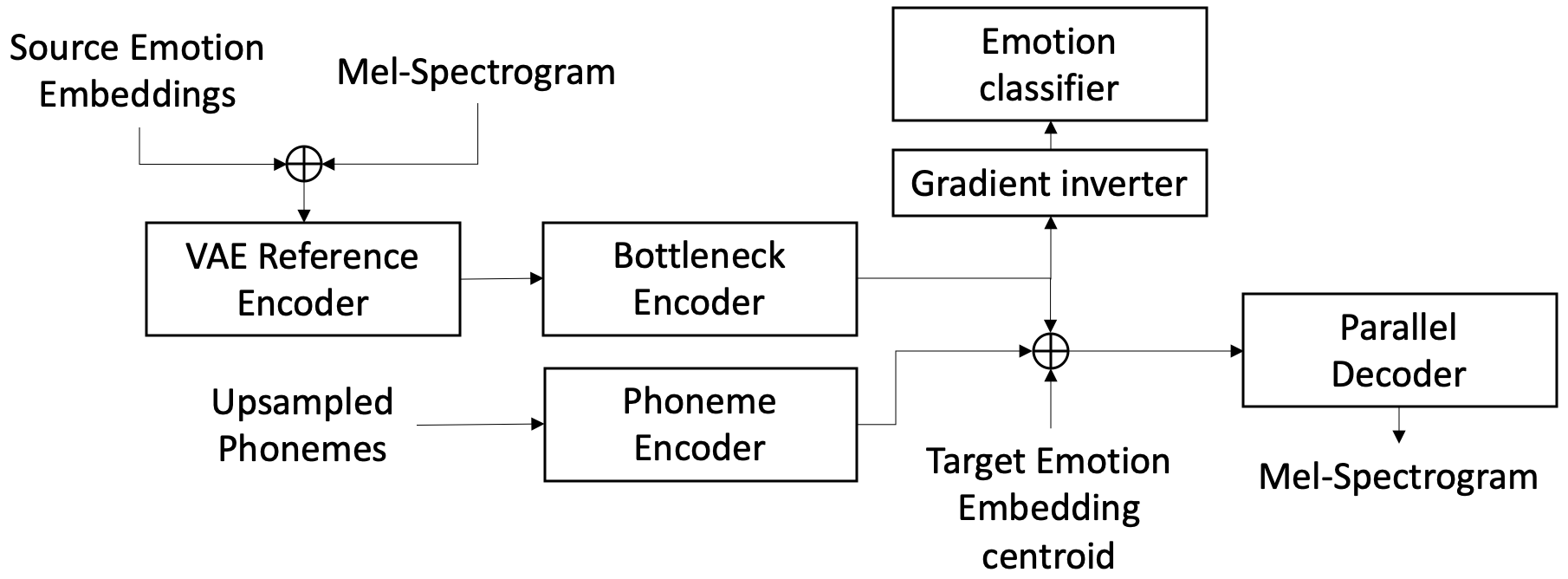}
	\caption{Structure of the encoder-decoder EmoCat model with a gradient inverter block followed by an emotion classifier to remove emotion information in the bottleneck embeddings. The plus sign denotes a concatenation.
	}
	\label{fig:EmoCat}
\end{figure}

In this section we introduce EmoCat, a language-agnostic intra-speaker emotion conversion model. It aims to convert neutral speech to emotional speech of the same speaker\footnote{We have informally verified that it also allows conversion between emotions, but this lies out of the main scope of this paper.}. 
EmoCat is based on CopyCat \cite{karlapati2020copycat} and inherits the same structure and hyper-parameters except four differences:
\begin{enumerate}
	\item It uses 64-dim emotion embeddings instead of 128-dim speaker embeddings (see Section \ref{ssec:emo_embeddings}).
	\item It uses a gradient inverter block to remove emotion leakage from the bottleneck embeddings (see Section \ref{ssec:emotion_leakage_removal}).
	\item It operates on multi-lingual data (see Section \ref{ssec:database}).
	\item It does not pass the phoneme embeddings to the VAE reference encoder.
\end{enumerate}
Figure \ref{fig:EmoCat} shows the network structure. The VAE reference encoder encodes the mel-spectrograms and its emotion embedding. A dimensional and temporal bottleneck is applied by only selecting every N-th frame \cite{qian2019autovc}. Each selected frame is copied N times (to restore the sequence length). The bottleneck embeddings should contain as much information as possible to generate high-quality speech but no emotion information. This is ensured by passing them through the gradient inverter to the emotion classifier, which removes any leaking emotion information. Force-aligned upsampled phonemes (procedure described in \cite{klimkov2019fine}) are encoded by the phoneme encoder to produce phoneme embeddings. During inference, the bottleneck and phoneme embeddings are stacked with the target-emotion em- bedding centroid and consumed by a parallel decoder to produce the converted mel-spectrograms. During training, the oracle utterance- level emotion embedding is used on the encoder and decoder side. Source and target spectrograms are the same as well. The parallel decoder consists of a stack of three convolutional layers followed by a uni-directional long short-term memory (LSTM). The model is trained with an L1 reconstruction loss and the KL-loss on the VAE latent space. For the detailed architecture, please refer to the original CopyCat paper \cite{karlapati2020copycat}.

\vspace{\sectioncut}
\subsection{Utterance-level emotion embeddings}
\label{ssec:emo_embeddings}
During training, utterance-level emotion embeddings are fed to the VAE reference encoder and the parallel decoder. 
The emotion embeddings need to be organised language-independently by their style and other latent information to be beneficial to the model. This excludes simple embeddings per emotion class and suggests a learn-able approach. Thus we obtain them from a Tacotron-like TTS model \cite{latorre2019effect} with the addition of two VAE reference encoders \cite{tyagi2019dynamic}. One reference encoder captures the speaker information while the other captures the emotion. We use intercross training \cite{bian2019multi} to guide each encoder to encode only the speaker/emotion information and to be language-independent. We use the predicted embeddings from the emotion reference encoder as utterance-level emotion embeddings for the EmoCat training. We could learn the emotion embeddings in a similar fashion on-the-fly within the EmoCat model, but this would increase its training time, which is not desirable during research. We could also obtain them from a simple emotion recognition model, but we hypothesised that those embeddings might be more suited for recognition than generation.

For the CopyCat model, robust speaker embeddings from a pre-trained speaker identification system are necessary, because the model also has to convert from unseen speakers. This is not the case for the EmoCat model, which only converts between seen emotions. Thus it requires less sophisticated emotion embeddings.

During inference, the utterance-level emotion embedding of the converted spectrograms is unknown. Instead we compute the centroid for each emotion over all emotion embeddings extracted from the training set and feed it to the decoder. The VAE reference encoder still uses the utterance-level emotion embedding of the input audio.

\vspace{\sectioncut}
\subsection{Gradient inverter}
\label{ssec:emotion_leakage_removal}

As emotion is a continuous and integral part of speech, it is necessary to explicitly prevent it from leaking from the encoder to the decoder side.
With a pre-trained EmoCat with frozen weights we trained independent gated recurrent unit (GRU) emotion classifiers to predict the source emotion from the bottleneck embeddings, where the best achieved 64\% overall accuracy. We found that heavy leakage resulted in low emotion intensity during conversion.
Decreasing the bottleneck (as described in AutoVC \cite{qian2019autovc}) led to heavy degradation in signal quality and intelligibility. With the reconstruction loss alone, we could not force the bottleneck embeddings to remove the undesired emotion information while keeping information needed for high signal quality.

Instead we used a gradient reversal block before the emotion classifier during training to actively remove emotion leakage from the bottleneck embeddings. 
The idea of gradient reversal is to reverse the gradients during back-propagation to remove any activation in the input that helps the following classifier. Gradient reversal achieves this by swapping the sign of the gradient $\Delta$ (Equation \ref{eq:grad_reversal}). It also applies a weight $\lambda$ to control the impact of the gradient on the preceding layers. The choice of the weight greatly influences the performance of the final model. 

\begin{equation}
\Delta' = -\lambda \Delta
\label{eq:grad_reversal}
\end{equation}

We experimented with a feed-forward and a GRU based emotion classifier. Interestingly, EmoCat converged to a better model in terms of conversion ability with the feed-forward classifier than the GRU one. This suggests that with gradient reversal even a weak classifier gives sufficient gradients to lead to a better convergence point.

We again trained the same emotion classifier as above on the bottleneck embeddings of the model with gradient reversal. The classifier mainly predicted the majority class (95\% of the time) showing that the majority of the emotion leakage was removed. Informal listening verified that the conversion ability of the model improved. 

We argue that a simple swap of the sign (Equation \ref{eq:grad_reversal}) fulfils only half of the reversal purpose. Consider the following two scenarios:
\begin{enumerate}
	\item Imagine there is \textbf{no leakage} in the input. As the classifier cannot rely on any information in the input, its prediction is random and the cross-entropy loss on its predictions is high. Thus the back-propagated gradients are large as well. Even though there is no leakage the preceding network receives a large reversed gradient.
	\item Imagine there is \textbf{significant leakage} in the input and the classifier is already properly trained. Then its prediction is good, the cross-entropy loss is low, and the back-propagated gradients are small. Even though there is significant leakage the preceding network receives only a small reversed gradient.
\end{enumerate}
The desired effect on the preceding network in both scenarios should be swapped. Without any leakage the received gradients should be small, while with significant leakage the gradients should be large. 

To address this issue, we present the \textbf{gradient inverter} block. Instead of only swapping the sign of the gradient, it performs a proper inversion by also converting small gradients to large ones and vice versa. We have experimented with two gradient inverter functions.

\begin{align}
	\Delta' =& \frac{-\lambda \Delta}{||\Delta||^2_2}\quad&\text{Inverse square norm}\label{eq:squarenorm}\\
	\Delta' =& \frac{-\lambda \Delta}{\exp ||\Delta||^2_2}\quad&\text{Inverse exp square norm}\label{eq:expsquarenorm}
\end{align}

Equation \ref{eq:squarenorm} implements directly what we want to achieve by scaling the gradient by its squared norm. Gradients with a norm smaller than one will become greater than one and vice versa. However, it might lead to unstable behaviour as gradients with a norm close to zero are scaled towards infinity. Equations \ref{eq:expsquarenorm} prevents this by bounding the denominator to less than one. In this variant, gradients with a small norm remain almost unchanged while big gradients are quickly faded out. We found that depending on the target emotion one of the proposed inverter functions performs better.

\vspace{\sectioncut}
\subsection{Fine-tuning}
While the EmoCat model with the proposed gradient inverter achieved high emotion intensities, its signal quality left room for improvement.
\label{sssec:data_fine_tuning}
We investigated fine-tuning on a subset of the training data. First the model was trained with all data until convergence. Then we continued training on emotional and similar amounts of neutral data. This should compensate the averaging effect in the decoder introduced by the huge amount of neutral training data. We did not change any hyper-parameters, learning rates, or losses compared to the first training step.
This approach outperforms a GAN-like loss (same as used for CopyCat \cite{karlapati2020copycat}), which strives for the generated spectrogram to be indistinguishable from the recordings.

\vspace{\sectioncut}
\section{Experiments}
\label{sec:experiments}
\begin{figure*}[t]
	\centering
	\includegraphics[width=\textwidth]{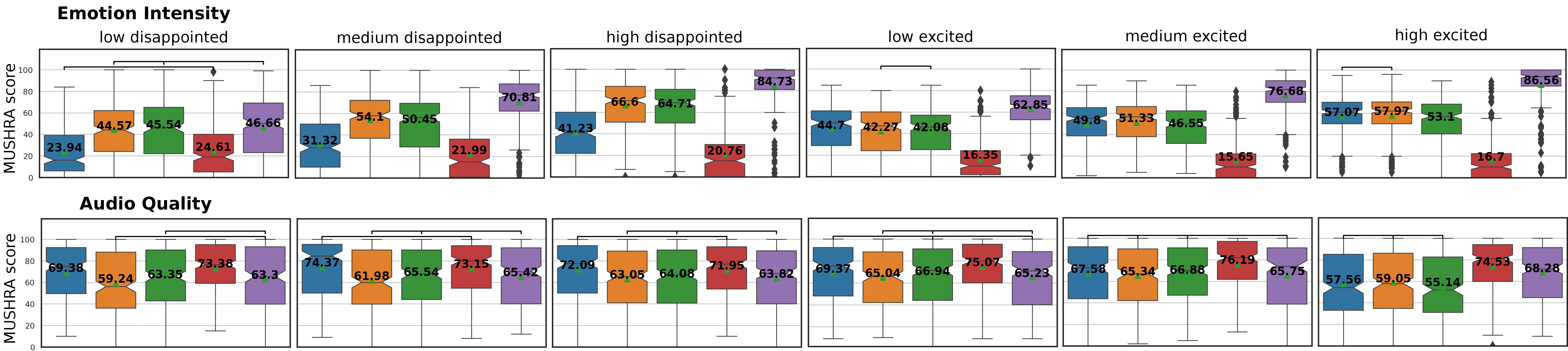}
	\caption{System descriptions: blue: grad. reversal, orange: grad. inverter, green: grad. inverter fine-tuned, red: neutral baseline, purple: recordings. Black horizontal bars connecting systems denote no statistically significant difference between them (\textit{p-value} $< 0.05$).}
	\label{fig:eval}
\end{figure*}

We aim at generating emotional German samples by converting from neutral using a model trained with a limited amount of emotional German data. We focused on two emotions: excited and disappointed, in three intensities: low, medium, high.

\vspace{\sectioncut}
\subsection{Database}
\label{ssec:database}
We use two internal databases. For German, we use more than \mbox{20 h} of neutral and \mbox{45 min} of emotional single-speaker recordings of a female voice. 20 neutral samples are set aside as test set. We do not use a development set to guide the training because the L1 reconstruction loss does not match human perception. The 45 min of emotional data are split equally into excited and disappointed. 25\% is low, 50\% medium, and 25\% high intensity. Excluding the test set, we have around 5 min for the most challenging intensity: high.
As we do not have access to more emotional German data, we use recordings of a female US English voice as supporting speaker. From this speaker, we use more than 20 h of neutral and more than 10 h of emotional recordings of the same emotion categories. We found that including US English data greatly improved the conversion abilities of our model, despite the differences in language. 24 kHz recordings are used. We trim all silences to be maximum 100 ms and extract 80-dim mel-spectrogram. We use phonemes with fully disjoint sets for English and German, thus the speaker identity can directly be inferred and explicit speaker embeddings are unnecessary.

\vspace{\sectioncut}
\subsection{Models}
We conduct an ablation study across three models. Each is trained for 100k steps on the combined two databases. The mel-spectrogram is synthesised with our universal vocoder \cite{lorenzo2019towards}.
\begin{enumerate}
	\item \textbf{Grad. reversal} - This model uses the vanilla gradient reversal block (Equation \ref{eq:grad_reversal}) to remove leaking emotion information. In contrast to the following two models, we used a weighted cross entropy loss for the adversarial emotion classifier to compensate for the huge class imbalance in the training data. We chose the weights inverse proportional to the amount of the emotion in the total training data. We found that this improved the grad. reversal model.
	\item \textbf{Grad. inverter} - This model replaces the gradient reversal block of model 1 with the improved gradient inverter block (see Section \ref{ssec:emotion_leakage_removal}). We use two separate models for the conversion. The model to convert to the three excited emotions uses the inverse exp square norm function (Equation \ref{eq:expsquarenorm}), while the one to convert to disappointed uses inverse square norm (Equation \ref{eq:squarenorm}). This was selected based on a clear performance difference in informal listening.
	\item \textbf{Fine-tuning} - This is model 2 fine-tuned for 2k steps as described in Section \ref{sssec:data_fine_tuning}. The best results were obtained by fine-tuning on the emotional data of the target speaker with a similar amount of neutral data as for each emotion. The neutral data requirement is probably due to the adversarial training. This simple fine-tuning outperforms GAN fine-tuning.
\end{enumerate}
We wanted to include a state-of-the-art baseline, however we did not find any work on emotion conversion from spectrograms. We adapted the work of \cite{rizos2020stargan} based on their StarGAN implementation\footnote{https://github.com/glam-imperial/EmotionalConversionStarGAN} to use mel-spectrograms instead of WORLD vocoder features, but the quality of the synthesized speech was very low. It is likely that major adaptations to the model architecture are necessary to achieve competitive results. However, creating such a baseline system is out of scope for this work. Therefore it was impossible for us to include a competitive state-of-the-art baseline model in our benchmark.

\vspace{\sectioncut}
\subsection{Evaluations}
\label{sec:evaluations}
We randomly selected 10 neutral German samples from the held-out test set and converted them to each of the six emotion intensities. 24 native German listeners rated the samples in terms of emotion intensity and audio quality in a MUSHRA \cite{series2014method} test from 0 to 100.

\vspace{\sectioncut}
\subsubsection{Emotion intensity}
We asked listeners to rate the emotion intensity where we provided another neutral recording (different sentence) as a reference of 0. We also included another recoding of the same emotion of the target speaker as an upper anchor and the utterance generated by a neutral baseline system. We see in Figure \ref{fig:eval} top line that our gradient inverter model outperforms vanilla gradient reversal for medium excited and is similar in high excited (no statistical difference, two-tailed t-test with \textit{p-value} $< 0.05$, denoted as a horizontal bar in the plots) while it is significantly worse for low excited. The exp square norm function (Equation \ref{eq:expsquarenorm}) only scales large gradients down which does not seem to be optimal for the excited intensities. For disappointed the gradient inverter model achieves more than 20 MUSHRA points higher score across all intensities, proving the improvement through the gradient inverter function. We either did not yet find a gradient inverter function which generalises to different emotions, or the function should be chosen depending on the use case.
Fine-tuning lowers the emotion intensity for the medium and high emotions. This shows an averaging effect of the neutral and low intensity data. It should also be noted that we see a clear ascent from the low to the high intensity, but do not yet reach the emotion intensity of the recordings except for low disappointed. We were only able to partially address the averaging effect in the decoder, which might reveal a general shortcoming of current decoder architectures. Highly expressive data in another language seems to improve the system only to a certain point. More high expressive German recordings, even from other speakers, might push the emotion intensity further.

\vspace{\sectioncut}
\subsubsection{Audio quality}

We compared the same systems as above but without a reference sample and asked the listeners to rate the audio quality (Figure \ref{fig:eval} bottom line). We do not see a statistical difference between all systems for medium and high excited.
Vanilla gradient reversal outperforms both other techniques for low excited and all disappointed intensities, but they are still at par with the recordings. We see a trade-off between emotion intensity and audio quality here. We usually found that higher emotion intensities suffer from reduced signal quality. Most likely because low intensities are close to neutral samples for which we have a lot of training data. This leads back to the averaging effect in the decoder. We suggest to explore different decoder architectures more suitable for highly expressive speaking styles.
While we are not able to reach the emotion intensity of the recordings yet, we achieve high audio quality at a generally lower intensity level.
Fine-tuning did not achieve the desired improvement in audio quality. Even though it increased the MUSHRA score in five out of six emotions the difference is only statistically significant for low disappointed. The increase in audio quality might be a consequence of the lower emotion intensity instead of fine-tuning. However, for low disappointed fine-tuning increased audio quality without reduced emotion intensity. 

\vspace{\sectioncut}
\section{Conclusion}
\label{sec:conclusion}
We proposed EmoCat, a novel language-agnostic EVC model based on CopyCat, which operates directly on mel-spectrograms. It allows to convert neutral to emotional samples in German with less than 45 minutes of German emotional recordings. It achieves this by leveraging large amounts of emotional English data with the same emotions. Even though the model is able to generate expressive speech at different intensities, we are not yet matching the expressiveness of the recordings. Moreover, we presented the gradient inverter block, an improvement to gradient reversal. This showed statistical significant improvements in emotion intensity for four out of six emotions in subjective listening tests. We also found minor improvements in audio quality, at the cost of emotion intensity, through fine-tuning on the target emotional data. Future work is required to investigate the influence of increasing the amount of emotional German data and further improvements to the gradient inverter functions.

\bibliographystyle{IEEEbib}
\bibliography{refs}

\end{document}